\documentstyle[graphicx, 12pt]{article}
\begin{document}

\small
\hoffset=-1truecm
\voffset=-2truecm
\title{\bf The circular loop equation of a cosmic string in Kerr-de Sitter spacetimes}
\author{Zhengyan Gu\footnote {E-mail address: guzhengyan@citiz.net} \hspace {1cm} Hongbo Cheng\footnote {E-mail address:
hbcheng@public4.sta.net.cn}\\
Department of Physics, East China University of Science and
Technology,\\ Shanghai 200237, China}

\date{}
\maketitle

\begin{abstract}
The equation of cosmic string loops in Kerr-de Sitter spacetimes
is derived. Having solved the equation numerically, we find that
the loops can evolve except for too small ones in the spacetimes.
\end{abstract}
\vspace{8cm}
\hspace{1cm}
PACS number(s):

\newpage
Cosmic strings including their formation, evolution and
observational effects attracted more attentions in the eighties
and much of the nineties [1, 2]. As linear defects at a symmetry
breaking phase transition, cosmic strings can be produced at the
end of an inflation. The cosmic strings have several cosmological
importance especially two features. One was an adequate
explanation for originating the primordial density perturbations
which are necessary for galaxies and clusters, and the other is
their spacetime metric with deficit angle. However, with expected
improvements in observational data, the CMB and WMAP experiments
favour the models without cosmic strings due to their weaker
tension $G\mu\leq10^{-6}$, which leads the strings not to seed the
large scale structure formation of the universe [3-5].

There has been a recent resurgence of interest in cosmological
consequences of cosmic strings for both theoretical and
observational reasons. The topological defects including cosmic
strings can be inevitably formed at the end of brane inflation [6,
7], which provides us with a potential window on the M theory on
the theoretical side [8-10]. On the experimental ones, there are
some examples to be found. Sazhin et al discovered a gravitational
lens called CSL-1 [11, 12] which is supposed to invoke two images
of comparable magnitude of the same giant elliptical galaxy. They
found a lot of similar objects in the vicinity. In addition,
Schild et al observed and analysed the anomalous brightness
fluctuations in a multiple-image lens system such as Q0957+561A, B
[13, 14], and the phenomena is interpreted as lensing by an
oscillating loops of cosmic string.

It is necessary to continue exploring the evolution of cosmic
string loops. Once the cosmic strings formed at any epoch in the
history of the universe, curve sections of strings would oscillate
under their own force of tension. Although the strings stretch
under the influence of the Hubble expansion and the oscillating
strings lose energy to gravitational radiation, they collide and
intersect to undergo reconnections. The reconnections of long
strings and large loops will produce small loops copiously. In
general the string networks consist of long strings and closed
string loops. The cosmic string loops oscillate with time rather
randomly. They are thought as complicated time dependent
gravitational source. The gravitationally lensed quasar
Q0957+561A, B has been studied intensively for 25 years [13, 14].
Schild et al put forward an analysis of brightness fluctuations in
the system consisting of two quasar images separated by
approximately $6''$. They are known to be images of the same
quasar not only because of the spectroscopic match, but also
because the images fluctuate in brightness, and the time delay
between fluctuations is always the same. They further suggested
that the effect may be due to lensing by an oscillating loop of
cosmic string between us and the lensing system because loops of
cosmic strings supply quantitative explanations of synchronous
variations in the two images of the gravitationally lensed quasar
Q0957+561A, B. Therefore the important results such as synchronous
variations in the two images with no time delay may be due to the
oscillating loops of cosmic string [14]. In a word, the
theoretical and observational results support the existence of
cosmic string loops. The evolution and fate of cosmic string loops
also attracted more attention [1, 2, 15-21]. In some cases like
Minkowski space and Robertson-Walker universe the loops will
collapse to form black holes or become a long cosmic string
instead of keeping the oscillating loops. It was showed that the
loops of cosmic string in de Sitter spacetimes will keeps up
expanding if their initial radii are large enough [19]. That a
loop with smaller initial radius may expand in some spacetimes is
supported by the observational evidence mentioned above. Clearly
it is necessary to investigate the loops in some other spacetimes
in order to explore the possibility that the smaller loops of
cosmic string can live.

The Kerr-de Sitter spacetimes are currently of great interest. It
is a remarkable fact that there are roughly $10^{20}$ rotating
black holes in the observable universe, so the surrounding of
astrophysically relevant black hole can be described by the Kerr
metric [22]. Recently, this kind of spacetime is popular because
of its distinct features [23-27]. The Kerr spacetime is stable
against massless field perturbations, but it is unstable for
massive case because of superradiance. The phenomenon is that the
energy of the reflected wave is larger than the incident one in a
scattering process. According to the observational evidence that
the universe expands faster than we thought [28, 29], it is worth
investigating the so called Kerr-de Sitter spacetime describing
rotating gravitational sources with a positive cosmological
constant. It is necessary to study the Kerr-de Sitter metrics in
various directions.

There should be more cosmic string loops existing in our universe
in order to explain the astronomical phenomena [13, 14]. However a
large loop will evolve to be a lot of smaller loops and the tiny
loops can not inhabit in de Sitter background unless they are very
large. The purpose of this paper is to derive the equation of
circular loops of cosmic string in the Kerr-de Sitter spacetime.
After the loops formation, the circular loops contract under their
own tension to form black holes inevitably in the Minkowski,
Robertson-Walker universe. In the de Sitter backgrounds, only
loops with large initial radii can avoid becoming black holes
[19]. It is interesting that for the Kerr-de Sitter cases, the
circular loops with initial radius which is much smaller than that
in de Sitter spacetime will expand instead of going towards the
rotating gravitational source. At first we derive the equations of
circular loops of cosmic string in Kerr-de Sitter spacetimes. We
solve the equations numerically. The conclusions are emphasized at
last.

We start to consider the evolution of cosmic strings in a Kerr-de
Sitter spacetime. The metric describing the environment is written
as,

\begin{equation}
ds^{2}=\frac{\Delta_{r}}{\rho^{2}}(dt-\frac{a}{\Xi}\sin^{2}\theta
d\phi)^{2}-\frac{\rho^{2}}{\Delta_{r}}dr^{2}-\frac{\rho^{2}}{\Delta_{\theta}}d\theta^{2}
-\frac{\Delta_{\theta}\sin^{2}\theta}{\rho^{2}}(adt-\frac{r^{2}+a^{2}}{\Xi}d\phi)^{2}
\end{equation}

\noindent where

\begin{equation}
\Delta_{r}=(r^{2}+a^{2})(1-\frac{r^{2}}{L^{2}})-2mr
\end{equation}

\begin{equation}
\Delta_{\theta}=1+\frac{a^{2}}{L^{2}}\cos^{2}\theta
\end{equation}

\begin{equation}
\rho^{2}=r^{2}+a^{2}\cos^{2}\theta
\end{equation}

\begin{equation}
\Xi=1+\frac{a^{2}}{L^{2}}
\end{equation}

\noindent and $L=\sqrt{\frac{3}{\Lambda}}$ is the de Sitter radius
associated with the cosmological constant $\Lambda$. The
gravitational source has mass $m$, angular momentum $J=ma$, and an
event horizon at $r=r_{+}$, the largest root of $\Delta_{r}$. A
free string propagating in a spacetime sweeps out a world sheet
which is two-dimensional surface. The Nambu-Goto action is used to
describe the motion of string and is given by,

\begin{equation}
S=-\mu\int d^{2}\sigma[(\frac{\partial
x}{\partial\sigma^{0}}\cdot\frac{\partial
x}{\partial\sigma^{1}})^{2}-(\frac{\partial
x}{\partial\sigma^{0}})^{2}(\frac{\partial
x}{\partial\sigma^{1}})^{2}]^{\frac{1}{2}}
\end{equation}

\noindent where $\mu$ is the string tension.
$\sigma^{a}=(\tau,\sigma)$ $(a=0,1)$ are timelike and spacelike
string coordinates respectively. $x^{\mu}(\tau,\sigma)$
$(\mu,\nu=0,1,2,3)$ are the coordinates of the string world sheet
in the spacetime.

For simplicity we assume that the string lies in the hypersurface
$\theta=\frac{\pi}{2}$, then the spacetime coordinates of the
world-sheet parametrized by $\sigma^{0}=t$, $\sigma^{1}=\varphi$
can be chosen as,

\begin{equation}
x=(ct, r(t, \varphi), \frac{\pi}{2}, \varphi)
\end{equation}

In the case of planar circular loops, we have $r=r(t)$. According
to the metric (1) and coordinates (7), the action (6) is reduced
to,

\begin{equation}
S=-\mu\int\int dt d\varphi\sqrt{\Delta}
\end{equation}

\noindent which leads to the following equation of motion for
loops,

\begin{equation}
\frac{d}{dt}\frac{\partial\Delta}{\partial
\dot{r}}-\frac{1}{2\Delta}\frac{d\Delta}{dt}\frac{\partial\Delta}{\partial\dot{r}}
-\frac{\partial\Delta}{\partial r}=0
\end{equation}

\noindent where

\begin{equation}
\Delta=[\frac{a(r^{2}+a^{2})-a\Delta_{r}}{r^{2}\Xi}]^{2}
-\frac{(r^{2}+a^{2})^{2}-\Delta_{r}a^{2}}{\Delta_{r}\Xi^{2}}\dot{r}^{2}
+(\Delta_{r}-a^{2})\frac{(r^{2}+a^{2})^{2}-\Delta_{r}a^{2}}{r^{4}\Xi^{2}}
\end{equation}

\noindent According to equations (2-5), equation (9) becomes

\begin{eqnarray}
2\frac{(r^{2}+a^{2})^{2}-\Delta_{r}a^{2}}{\Delta_{r}\Xi^{2}}
\{1+\frac{1}{\Delta}\frac{(r^{2}+a^{2})^{2}-\Delta_{r}a^{2}}{\Delta_{r}\Xi^{2}}\dot{r}^{2}\}\ddot{r}\hspace{4.5cm}\nonumber\\
+\frac{2(r^{2}+a^{2})}{\Delta_{r}^{2}\Xi^{2}}[2(1+\frac{a^{2}}{L^{2}})r^{3}
+2a^{2}(1+\frac{a^{2}}{L^{2}})r-6mr^{2}+2ma^{2}]\dot{r}\hspace{2.8cm}\nonumber\\
-\frac{(r^{2}+a^{2})^{2}-\Delta_{r}a^{2}}{\Delta\Delta_{r}\Xi^{2}}
\{2a(2ar\dot{r}-a\frac{d\Delta_{r}}{dt})\frac{(r^{2}+a^{2})-\Delta_{r}}{r^{4}\Xi^{2}}\hspace{3.5cm}\nonumber\\
-4[a(r^{2}+a^{2})-\Delta_{r}]\frac{(r^{2}+a^{2})-\Delta_{r}}{r^{5}\Xi^{2}}a\dot{r}\hspace{6cm}\nonumber\\
-(\frac{1}{\Delta_{r}\Xi^{2}}[4(r^{2}+a^{2})r\dot{r}-\frac{d\Delta_{r}}{dt}a^{2}]
-\frac{(r^{2}+a^{2})^{2}-\Delta_{r}a^{2}}{\Delta_{r}^{2}\Xi^{2}}\frac{d\Delta_{r}}{dt})\dot{r}^{2}\hspace{2cm}\nonumber\\
+\frac{d\Delta_{r}}{dt}\frac{(r^{2}+a^{2})^{2}-\Delta_{r}a^{2}}{r^{4}\Xi^{2}}
+\frac{\Delta_{r}-a^{2}}{r^{4}\Xi^{2}}[4(r^{2}+a^{2})r\dot{r}-\frac{d\Delta_{r}}{dt}a^{2}]\hspace{2cm}\nonumber\\
-4(\Delta_{r}-a^{2})\frac{(r^{2}+a^{2})^{2}-\Delta_{r}a^{2}}{r^{5}\Xi^{2}}\dot{r}\}\hspace{6cm}\nonumber\\
+\frac{4a^{2}}{r^{3}\Xi^{2}}(\frac{r^{3}}{L^{2}}+\frac{a^{2}}{L^{2}}r+2m)\hspace{7cm}\nonumber\\
-\frac{r^{2}+a^{2}}{\Delta_{r}^{2}\Xi^{2}}[2(1+\frac{a^{2}}{L^{2}})r^{3}
+2a^{2}(1+\frac{a^{2}}{L^{2}})r-6mr^{2}+2ma^{2}]\dot{r}^{2}\hspace{2.5cm}\nonumber\\
+\frac{1}{r^{3}\Xi^{2}}[-\frac{4r^{3}}{L^{2}}+2(1-\frac{a^{2}}{L^{2}})r-2m]
[(1+\frac{a^{2}}{L^{2}})r^{3}+(1+\frac{a^{2}}{L^{2}})a^{2}r+2ma^{2}]\hspace{0.5cm}\nonumber\\
+\frac{1}{r^{3}\Xi^{2}}[\frac{r^{3}}{L^{2}}-(1-\frac{a^{2}}{L^{2}})r+2m]
[2(1+\frac{a^{2}}{L^{2}})a^{2}r+6ma^{2}]=0
\end{eqnarray}

\noindent where

\begin{equation}
\frac{d\Delta_{r}}{dt}=[-\frac{4r^{3}}{L^{2}}+2(1-\frac{a^{2}}{L^{2}})r-2m]\dot{r}
\end{equation}

\noindent here the functions $\Delta_{r}$, $\Xi$ are denoted in
(2) and (5) respectively.

We shall consider the solution with $\dot{r}(0)=0$, and we assume
that loops are static when $t=0$, the time of formation of loops,
by neglecting the peculiar velocities of loops. For the circular
loops in de Sitter spacetimes, A. L. Larsen showed that the loops
with larger initial radius like $r(0)>0.707L$ expand instead of
becoming the black hole [19]. If the initial radius $r(0)<0.707L$,
the circular loop will collapse to form black hole in that
spacetime. Here $L$ is the de Sitter size. It was also pointed out
that the cosmic string loops aligned perpendicular to the spin
axis of a Kerr source will contract to be captured by the
gravitational source at last. We solve the equation of motion (11)
numerically by a Runge-Kutta algorithm to research on the
evolution of cosmic string loops in the Kerr-de Sitter spacetimes.
The evolution curves $\frac{r(t)}{L}=R(t)$ are plotted in figure 1
versus the different $\frac{r(0)}{L}=R(0)$, $0<R(0)<1$. The
numerical results show that there exist special parameter $R_{0}$
for a fixed $m$ and a fixed relative angular momentum denoted as
$A=\frac{a}{L}$. If $R(0)>R_{0}$, the loop will expand or they
will approach to the gravitational source instead of surviving.
The influence from source mass $m$ has been studied [19]. The
relation between the special parameter $R_{0}$ and $A=\frac{a}{L}$
for a fixed parameter like $M=\frac{m}{L}$ is shown in figure 2.
The special parameter $R_{0}$ is decreasing when the relative
angular momentum $A$ is increasing. Therefore, the influence of
rotating gravitational source, represented by parameter $A$, gives
rise to decreasing of initial radius of cosmic string loops. Thus,
around the rotating gravitational source in the presence of
positive cosmological constant, the smaller cosmic string loops
can become large to evolove instead of contracting to approach to
the source. It is interesting for us to find that the required
initial radii which lead the loops exist not to move towards the
source in Kerr-de Sitter backgrounds can be smaller than those in
de Sitter spacetimes.

According to the observational results, there should exist more
cosmic string loops in our Universe [13, 14]. Here we report the
evolution of planar circular loops of cosmic string in the
surrounding of a rotating body with positive cosmological constant
to investigate the fate of loops. The loops with relatively
smaller radius at the beginning  of formation expand and will not
collapse in the environment. With colliding and intersecting of
long strings and large loops, a lot of tiny loops are generated
continuously. In the de Sitter world most of the loops become
black holes [19], which means that few loops can survive. Our
findings on the evolution and existence of small loops indicate
that there may be a considerable number of loops of cosmic string
in our Universe due to a numerous number of rotating black holes
and positive cosmological constant. The same process that long
strings and large loops change into more small loops and most of
them expand and intersect into small loops go through again and
again, which can keep us observing the effect of cosmic string
loops.

The main result of this paper is equation(11), the circular loop
equation for a cosmic string evolving in the hypersurface with
$\theta=\frac{\pi}{2}$ in the Kerr-de Sitter spacetime. A
remarkable solution to this equation is that a loop may never
contract towards the source if its initial size is larger than a
special parameter denoted as $R_{0}$. Figure 2 shows that the
special parameter depends on the angular momentum of the rotating
gravitational source, larger the angular momentum, smaller the
special parameter. According to figure 2, all special parameters
for cosmic string loops around rotating gravitational sources with
positive cosmological constant are smaller than $0.707L$, the
special parameter for the loops in the de Sitter spacetime
obtained in [19]. Therefore a lot of cosmic string loops including
some smaller ones can evolve to survive in the Kerr-de Sitter
spacetime. The general evolution of cosmic string loops requires
further research.

\vspace{1cm}
\noindent \textbf{Acknowledge}

This work is supported by the Basic Theory Research Fund of East
China University of Science and Technology, grant No. YK0127312
and partly supported by the Shanghai Municipal Science and
Technology Commission No.04dz05905..

\newpage
\begin{figure}
\setlength{\belowcaptionskip}{10pt} \centering
  \includegraphics[width=15cm]{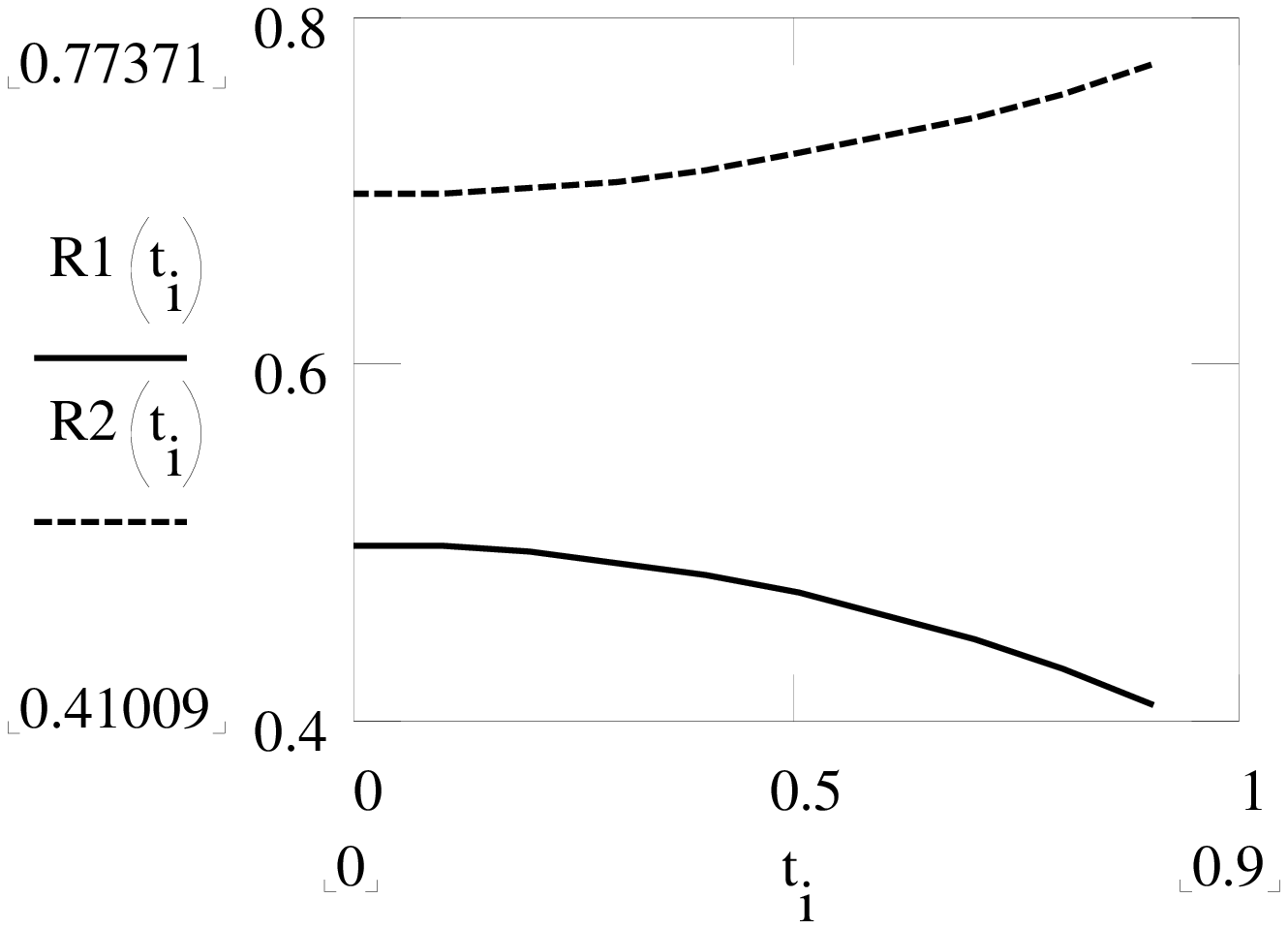}
  \caption{The evolution of loop size versus the different
  initial radii, solid line for $R(0)=0.5$ and dashed line for
$R(0)=0.7$}
\end{figure}

\newpage
\begin{figure}
\setlength{\belowcaptionskip}{10pt} \centering
  \includegraphics[width=15cm]{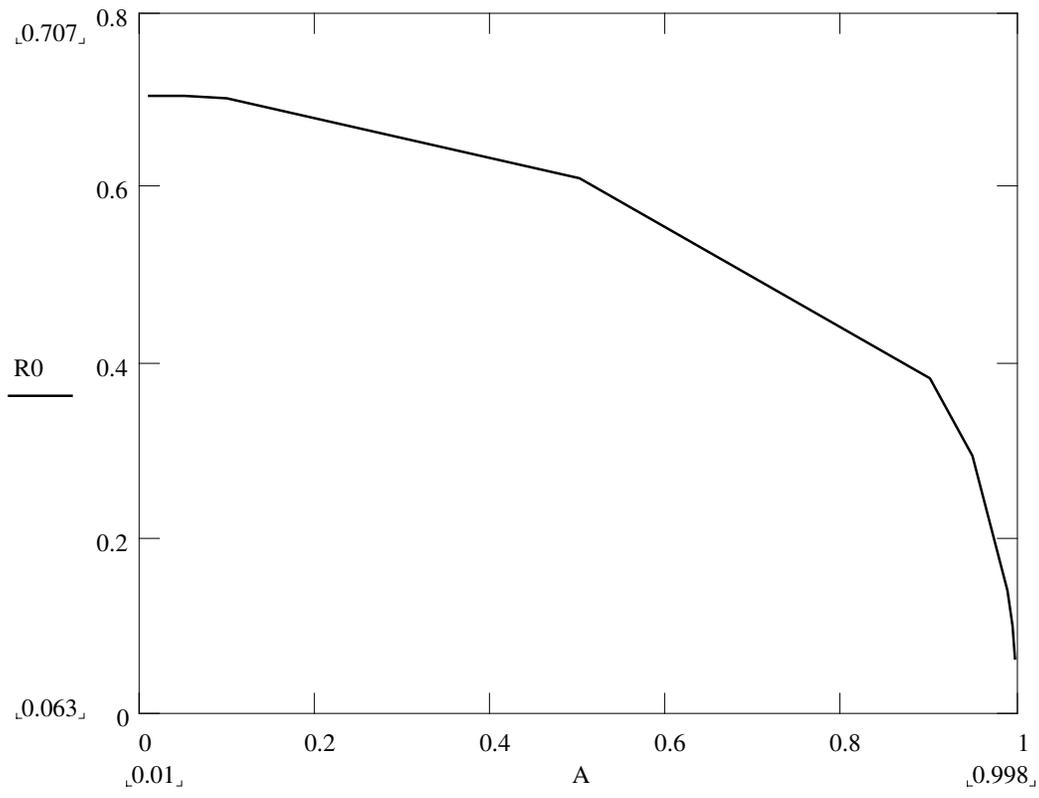}
  \caption{The special parameter $R_{0}$ as a function of
  the relative angular momentum $A$ for $\frac{m}{L}=0.0001$}
\end{figure}

\end{document}